# High resolution stopwatch for cents

### Z Gingl[1] and K Kopasz[2]

[1]Department of Technical Informatics, University of Szeged, Árpád tér 2, 6720 Szeged, Hungary

[2]Department of Experimental Physics, University of Szeged, Dóm tér 9, 6720 Szeged, Hungary

e-mail: gingl@physx.u-szeged.hu

**Abstract**
A very low-cost, easy-to-make stopwatch is presented to support various experiments in mechanics. The high-resolution stopwatch is based on two photodetectors connected directly to the microphone input of the sound card. A dedicated free open-source software has been developed and made available to download. The efficiency is demonstrated by a free fall experiment.

Recently, a simple stopwatch has been presented as a low cost tool to support experimentation in mechanics [1]. The buttons of a USB mouse have been replaced by photoresistors (aka photocells or light-dependent resistors); this way one can use light interruptions caused by mechanical movements to generate the mouse clicks. A simple software can measure the time elapsed between these events.

The aim of this paper is to present an alternative approach that is even simpler, easier to make, while the accuracy is higher. The principle is very simple: phototransistors can be connected directly to the microphone input port of the PC without any additional components, because the required bias voltage is provided on the microphone connector [2]. In this arrangement the phototransistor works like a light controlled switch in very much the same way as in the mouse-based project. However, the microphone input accepts analogue input signals, therefore it is much more sensitive and the signal is sampled much faster, typically at 44100 samples per second. This means that significantly less light is enough to operate this photogate, while the time resolution is much higher—it falls below 0.1 ms. No laser pointers are needed, therefore the comfort and reliability of experimenting are not affected by quickly discharging batteries and the requirement of a quite accurate mounting. Robust, bulky flashlight or DC powered lamps (LED lamps, USB lamps) can be used, instead. Mains powered lamps may be good enough, but the light intensity has an oscillation at the double of the mains frequency, which may add some uncertainty. One can easily check the waveforms using a free audio recording software like the popular Audacity [3].

The question is how we can connect two phototransistors to the typically monaural microphone input to start and stop our timer. It is pretty simple: just connect the two phototransistors in series as shown in figure 1.

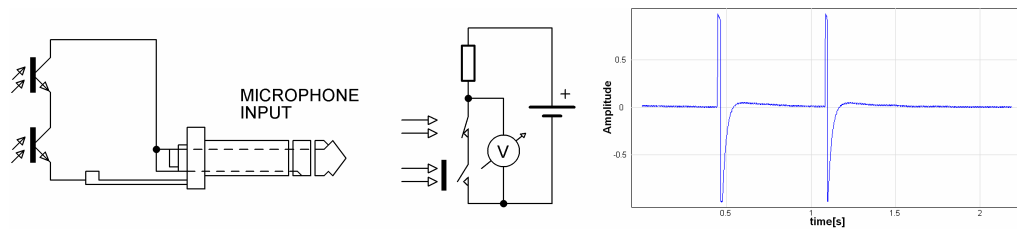

**Figure 1.** The microphone input of a PC can monitor two serially connected phototransistors. A simplified equivalent circuit is shown in the middle. If the light is blocked from any of the phototransistors, a voltage pulse will be measured at the microphone input. Pulses caused by two light interruptions are plotted on the right hand side.

If any of the phototransistors is blocked from the light beam, a voltage pulse will be generated at the microphone input that can be easily detected by software [4]. To ease mounting, the phototransistor can be fixed on a clothes peg, and even a simple ruler can be used to control the distance between the detectors quite accurately (see figure 2).

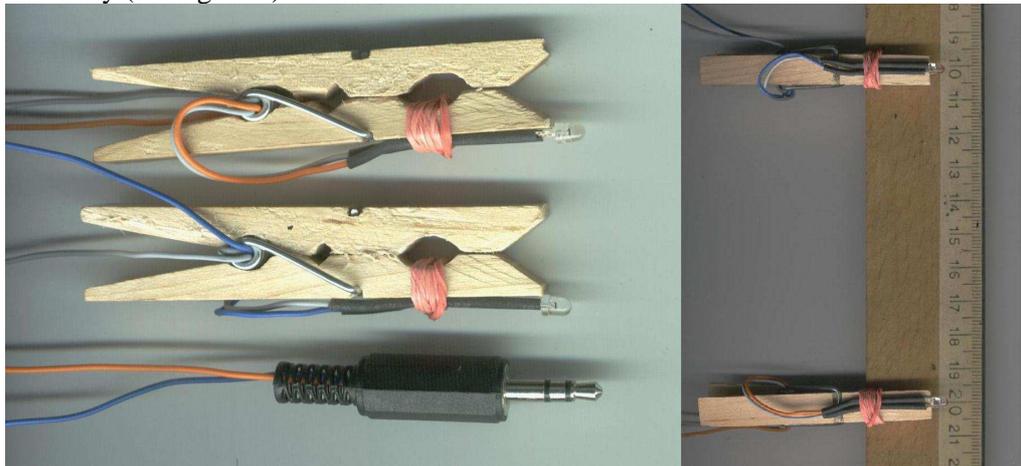

**Figure 2.** Two serially wired TEPT4400 phototransistors connected to the 3,5mm jack plug. Clothes-pegs can help to set the distance between the detectors easily and quickly.

The associated software is open-source and also available as a single executable [4]. It counts the number of events and keeps the history data that can be copied to the clipboard for further analysis in any spreadsheet software.

The performance of this stopwatch has been tested in a free fall experiment. A single lamp illuminated the two detectors from a distance of 0.5 m. The timer was started and stopped at 0.1 m and 0.2 m from the initial position, respectively. We have taken into account that beams from the light source reach the two detectors at different angles. The realisation of the experiment is illustrated in figure 3, while the arrangement and experimental data are shown in figure 4.

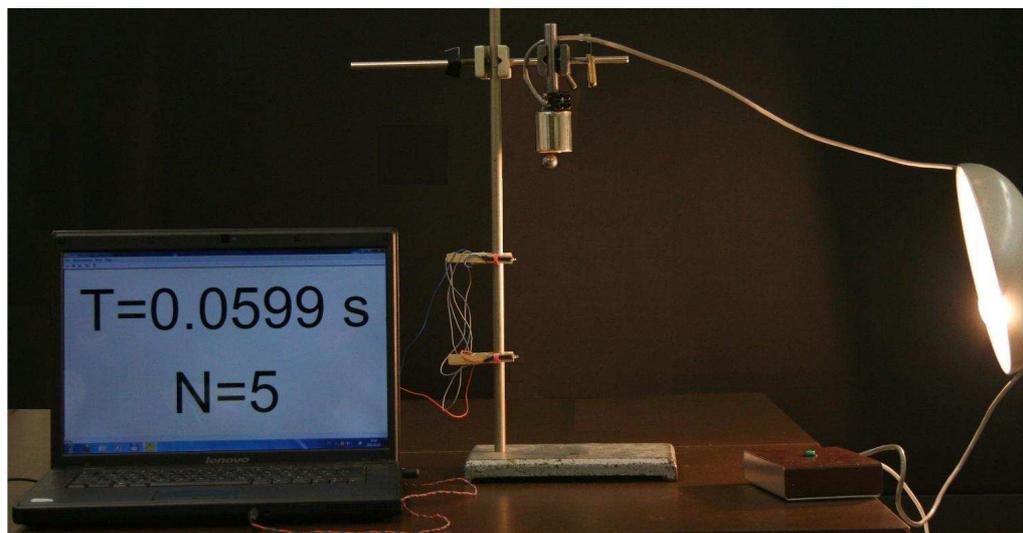

**Figure 3.** The setup of the free fall experiment. The serially connected phototransistors are plugged in the microphone port directly; an electromagnet is used

to drop the steel ball in front of the phototransistors. Although the mains lamp has quite high AC component, the sharp peaks generated when the ball blocks the light from the detector allow high accuracy event detection. The software displays the measured time and the number of detected events. All values are stored and can be copied into any spreadsheet software by 2 or 3 clicks.

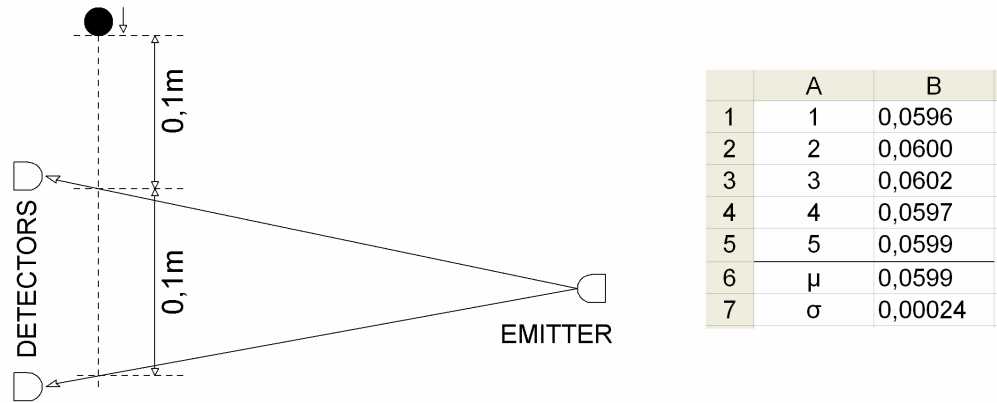

**Figure 4.** Free fall experiment demonstrates the high resolution and precision of the stopwatch. An Excel spreadsheet fragment shows the five measured values with mean ($\mu$=0.0599 s) and standard deviation ($\sigma$=0.00024 s).

Since only a jack plug (65 cents), two TEPT4400 phototransistors (72 cents/piece) and some wiring are needed, the overall cost is very low. Infrared phototransistors can be even cheaper: for example, the OFT-5301 is available for 10 cents [5]; but of course their use requires that the light source should contain intensive enough infrared components (LED lamps may not be sufficient). Note that even the mouse buttons(which in [1] were substituted with photoresistors whose function was to start and stop the measurement) can be replaced by phototransistors if the proper polarity is taken into account. Phototransitors outperform photoresistors in many parameters: they are faster, more sensitive and they have better spatial resolution.

It is not always emphasised that touching electronic components during operation is something to be certainly avoided. For isolation heat shrink tubes are the most elegant solution, but insulating tapes are also sufficient.

**Acknowledgment**

This work was supported by the Hungarian National Research Fund (OTKA K69018).

**References**

[1] Ganci A and Ganci S 2009 Stopwatch provides low-cost training *Phys. Educ.* **44** 119-20
[2] Gingl Z 2011 Sub-$10 photogate variants *Phys Teach. submitted*, Arxiv preprint *arXiv*:1101.0608
[3] http://sourceforge.net/projects/audacity/
[4] http://www.noise.physx.u-szeged.hu/edudev/stopwatch
[5] http://www.farnell.com, order codes: 1280662 (jack plug), 1497675 (phototransistor TEPT4400), 1716711 (phototransistor OFT-5301)